\documentclass[11pt,dvips]{article}
\usepackage{epsfig,times} 
\usepackage{picinpar}
\usepackage{wrapfig}
\usepackage{floatflt}
\makeatletter
\renewcommand{\section}{\@startsection{section}{1}{0in}
        {0.4\baselineskip}{0.1\baselineskip}{\Large\bf}}
\renewcommand{\subsection}{\@startsection{subsection}{2}{0in}
        {0.25\baselineskip}{-\baselineskip}{\large\bf}}
\renewcommand{\subsubsection}{\@startsection{subsubsection}{3}{0in}
        {0.1\baselineskip}{-\baselineskip}{\normalsize\bf}}
\makeatother
\def\lsim{\mathrel{\rlap{\lower4pt\hbox{\hskip1pt$\sim$}}
    \raise1pt\hbox{$<$}}}         
\def\gsim{\mathrel{\rlap{\lower4pt\hbox{\hskip1pt$\sim$}}
    \raise1pt\hbox{$>$}}}         
\newcommand{\incap}{\setlength{\baselineskip}{0.85\baselineskip}}
\newcommand{\href}[2]{#1}

\begin{document}

\setlength{\baselineskip}{0.94\baselineskip}
%
\thispagestyle{myheadings}
%
\markright{HE 5.1.05}
\begin{center}
%
{\LARGE \bf Cosmic antiprotons as a probe for neutralino dark matter?}
\end{center}

\begin{center}
%
%
{\bf L.~Bergstr{\"o}m$^{1}$, \underline{J.~Edsj{\"o}}$^{1*}$, 
and P.~Ullio$^{1}$}\\
{\it $^{1}$Department of Physics, Stockholm University, Box 6730, 
SE-113 85 Stockholm, Sweden\\
$^*$ presenter}
\end{center}

\begin{center}
{\large \bf Abstract\\}
\end{center}
\vspace{-0.5ex}
%
%
The flux of cosmic ray antiprotons from neutralino annihilations in
the galactic halo is computed for a large sample of models in the
Minimal Supersymmetric extension of the Standard Model.  We also
revisit the problem of estimating the background of low-energy cosmic
ray induced secondary antiprotons, taking into account their
subsequent interactions (and energy loss) and the presence of nuclei
in the interstellar matter.

We point out that in some cases the optimal kinetic energy to search
for a signal from supersymmetric dark matter is above several GeV,
rather than the traditional sub-GeV region. The large astrophysical
uncertainties involved do not allow the exclusion of any of the MSSM
models we consider, on the basis of current data.
%

\vspace{1ex}

%
%

\section{Introduction}

Among the most plausible candidates for the dark matter in the
Universe are Weakly Interacting Massive Particles (WIMPs), of which
the supersymmetric neutralino is a favourite candidate (see e.g.\
Jungman et al., 1997 for a review).  We will here consider the
antiproton flux from neutralino dark matter annihilating in the
galactic halo and we will also investigate the prospects of seeing
such a signal above the conventional background.

As antimatter seems not to exist in large quantities in the observable
Universe, including our own Galaxy, any contribution to the cosmic ray
generated antimatter flux (besides antiprotons also positrons) from
exotic sources may in principle be a good signature for such
sources. This issue has recently come into new focus thanks to
upcoming space experiments like {\sc Pamela} (Adriani et al.,
1995) and {\sc Ams} (Ahlen et al., 1994) with increased
sensitivity to the cosmic antimatter flux.

\section{Definition of the supersymmetric model}
\label{sec:MSSMdef}

We work in the minimal supersymmetric standard model with seven
phenomenological parameters and have generated about $10^5$ models by
scanning this parameter space (for details, see Bergstr{\"o}m et
al., 1999).  For each generated model, we check if it is excluded by
recent accelerator constraints of which the most important ones
are the LEP bounds (Carr, 1998) on the lightest chargino mass
(about 85--91 GeV), and the lightest Higgs boson mass $m_{H_{2}^{0}}$
(which range from 72.2--88.0 GeV) and the constraints from $b \to s
\gamma$ (Ammar et al., 1993 and Alam et al.,1995).

For each model allowed by current accelerator constraints we calculate
the relic density of neutralinos $\Omega_\chi h^2$ where the relic
density calculation is done as described in Edsj\"o and Gondolo
(1997), i.e.\ including so called coannihilations. We will only be
interested in models where neutralinos can be a major part of the dark
matter in the Universe, so we restrict ourselves to relic densities in
the range $0.025 < \Omega_\chi h^2 < 1$.

\section{Antiproton production by neutralino annihilation}

Neutralinos are Majorana fermions and will annihilate with each other
in the halo producing leptons, quarks, gluons, gauge bosons and Higgs
bosons. The quarks, gauge bosons and Higgs bosons will decay and/or
form jets that will give rise to antiprotons (and antineutrons which
decay shortly to antiprotons). 
The hadronization for all final states (including gluons) is simulated
with the well-known particle physics Lund Monte Carlo program {\sc
Pythia} 6.115 (Sj{\"o}strand, 1994).

To calculate the source function of $\bar{p}$ from neutralino
annihilation we also need to specify the halo profile. We 
we will here focus on the  modified isothermal distribution with a
local halo density of 0.3 GeV/cm$^3$.

\section{Propagation model and solar modulation}
\label{sec:prop}

We choose to describe the propagation of cosmic rays in the Galaxy by
a transport equation of the diffusion type as written by Ginzburg and
Syrovatskii (1964) (see also Berezinskii et al., 1990; Gaisser, 1990).

The propagation region is assumed to have a cylindrical symmetry: the
Galaxy is split into two parts, a disk of radius $R_h$ and height
$2\cdot h_g$, where most of the interstellar gas is confined, and a
halo of height $2\cdot h_h$ and the same radius. We assume that the
diffusion coefficient is isotropic with possibly two different values
in the disk and in the halo, reflecting the fact that in the disk
there may be a larger random component of the magnetic fields.
For the diffusion coefficient, we assume the same kind of rigidity
dependence as in Chardonnet et al.\ (1996) and  Bottino et al.\ (1998),
i.e.\ that $D(R) = D^0 \left(1+R/R_0\right)^{0.6}$.
As a boundary condition we assume that the cosmic rays can escape freely
at the border of the propagation region. For details about our
propagation model and how the solutions are obtained, see
Bergstr{\"o}m et al.\ (1999).

\begin{figure}[!t]
 \centerline{\epsfig{figure=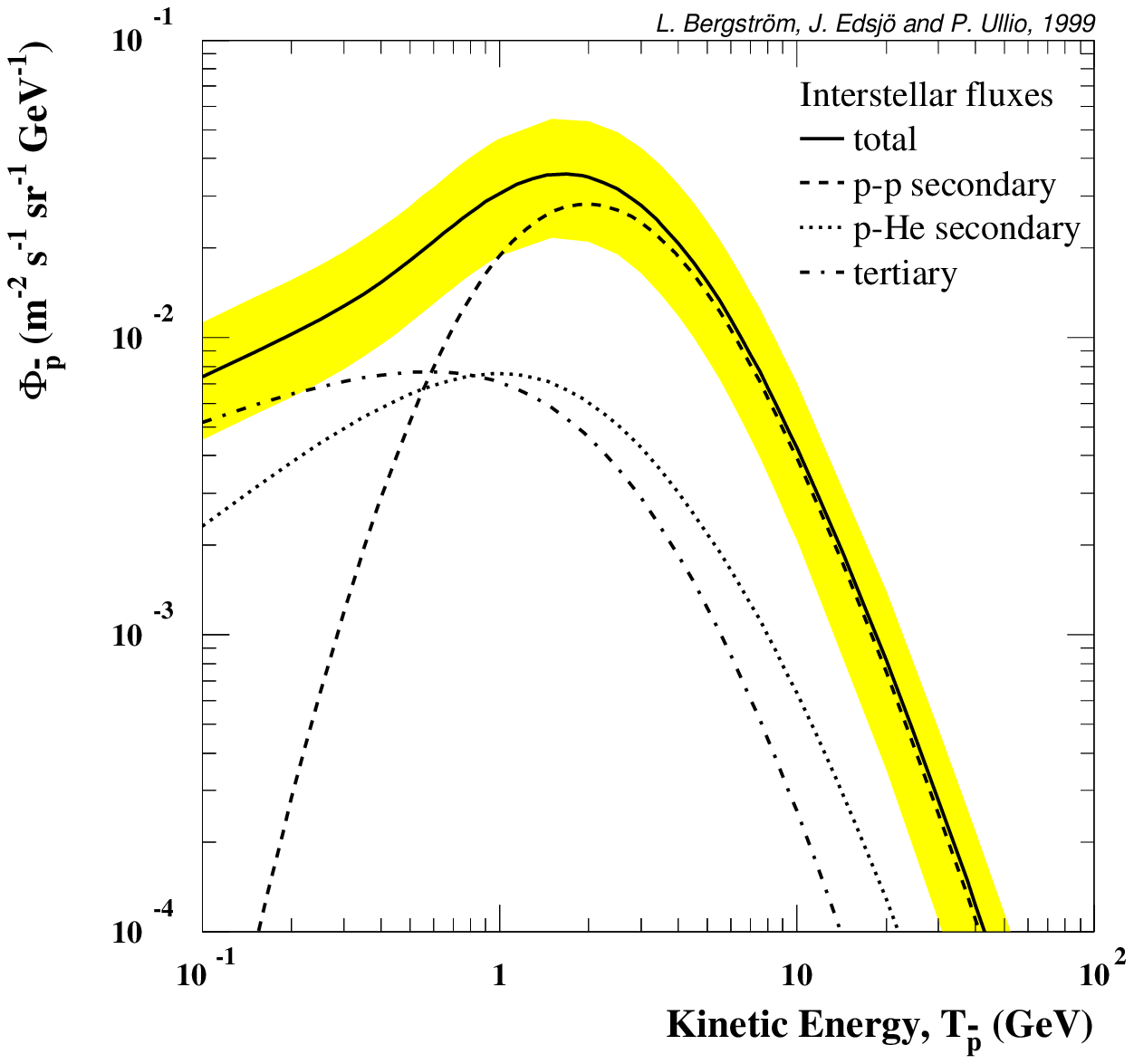,width=0.49\textwidth}
 \epsfig{file=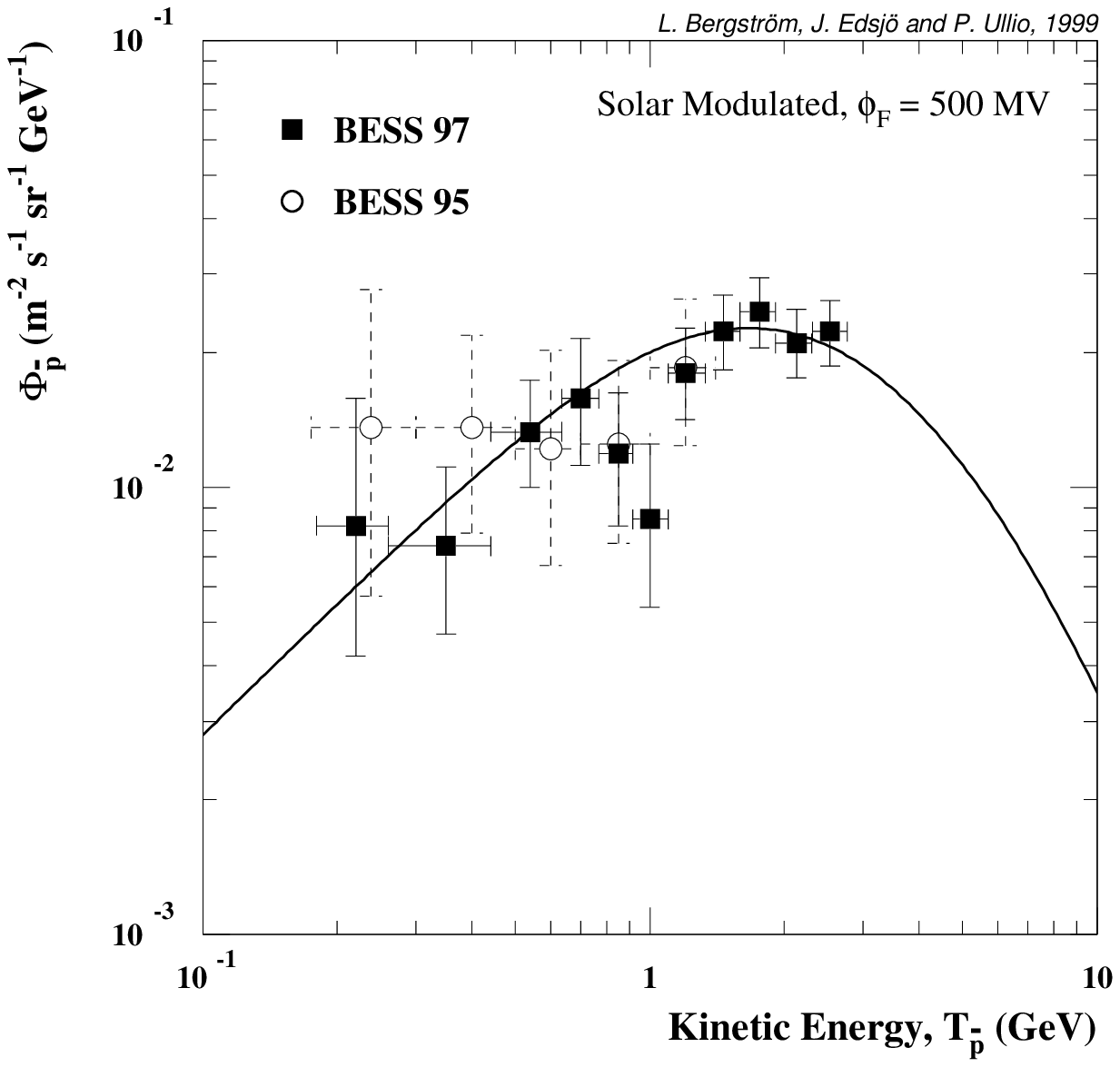,width=0.49\textwidth}}
 \caption[]{\incap a) The interstellar antiproton flux and the
   contribution from secondary and tertiary (i.e.\ $\bar{p}$s that 
   have lost energy) antiprotons.  The
   uncertainty due to the parametrization of the primary proton
   spectrum is also given as the shaded band.  b) The same as the
   solid line in a) but solar modulated with $\phi_F=500$ MV. The {\sc
     Bess} 95 and 97 data are also shown (Matsunaga et al., 1998;
   Orito, 1998).}
 \label{fig:back}
\end{figure}

For the solar modulation we use the analytical force-field 
approximation by Gleeson \& Axford (1967; 1968) for a spherically 
symmetric model.  To compare with the two sets of {\sc Bess} 
measurements, which are both near solar minimum, we choose the 
modulation parameter $\phi_F=500$~MV.

\section{Background estimates}

Secondary antiprotons are produced in cosmic ray collisions with the
interstellar gas. Normally, only $p-p$ interactions are included, 
which gives rise to a `window' at low energies with low fluxes.
However, we include $p-He$ interactions as well as $p-p$
interactions and also energy losses during propagation (with the full 
energy distribution).
Both of these processes tend to enhance the antiproton flux at low
energies and in Fig.~\ref{fig:back} (a) we show the background flux of
antiprotons and the contributions from $p-He$ interactions and energy
losses. We clearly see that the low-energy window has been filled-in. In
Fig.~\ref{fig:back} (b) we show the solar modulated curve compared
with recent {\sc Bess} measurements. We see that data is well
described by this conventional source alone.

\section{Signal from neutralino annihilation}

\begin{figure}[tb]
\centerline{\epsfig{file=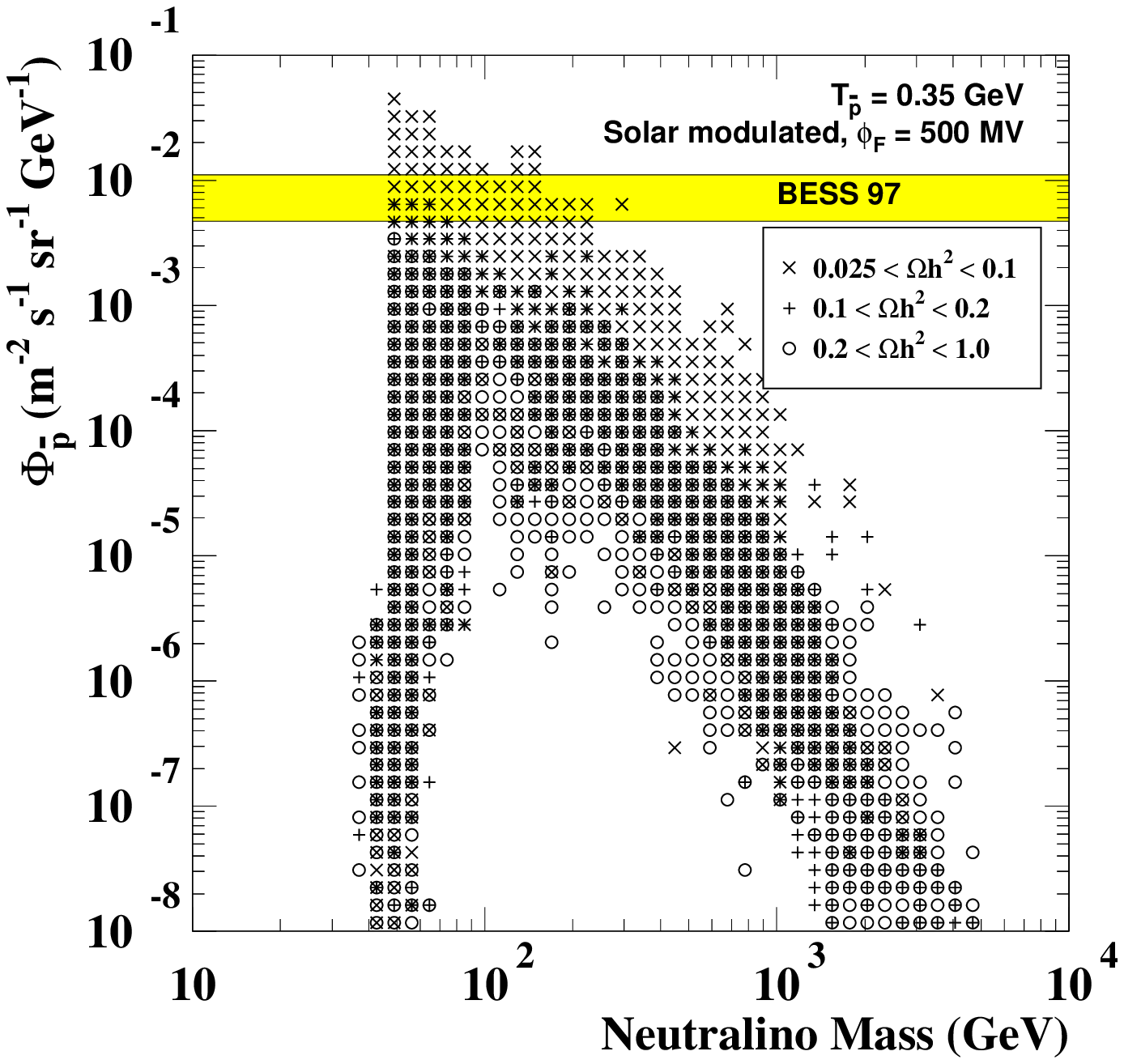,width=0.33\textwidth}
\epsfig{file=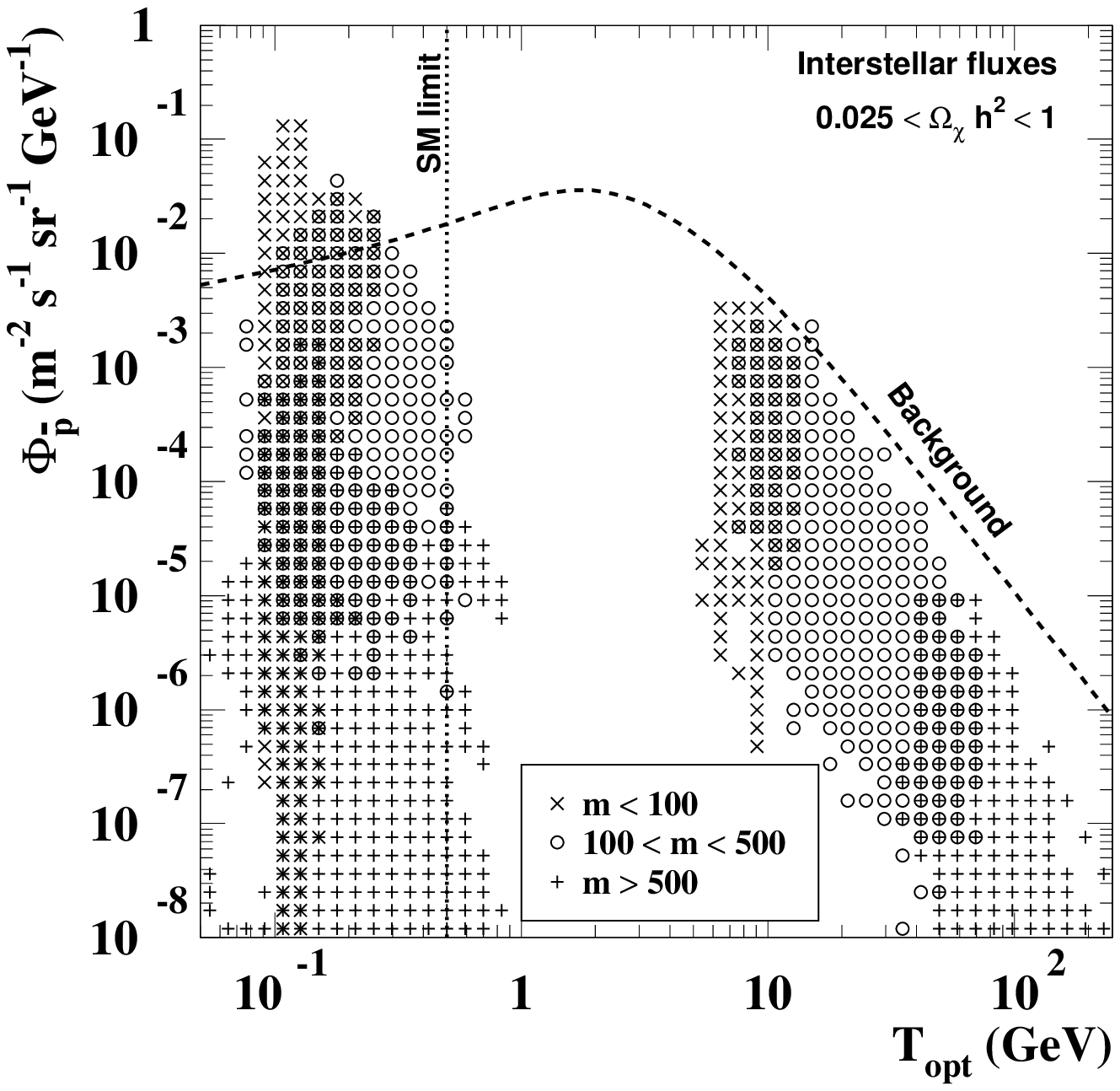,width=0.33\textwidth}
\epsfig{file=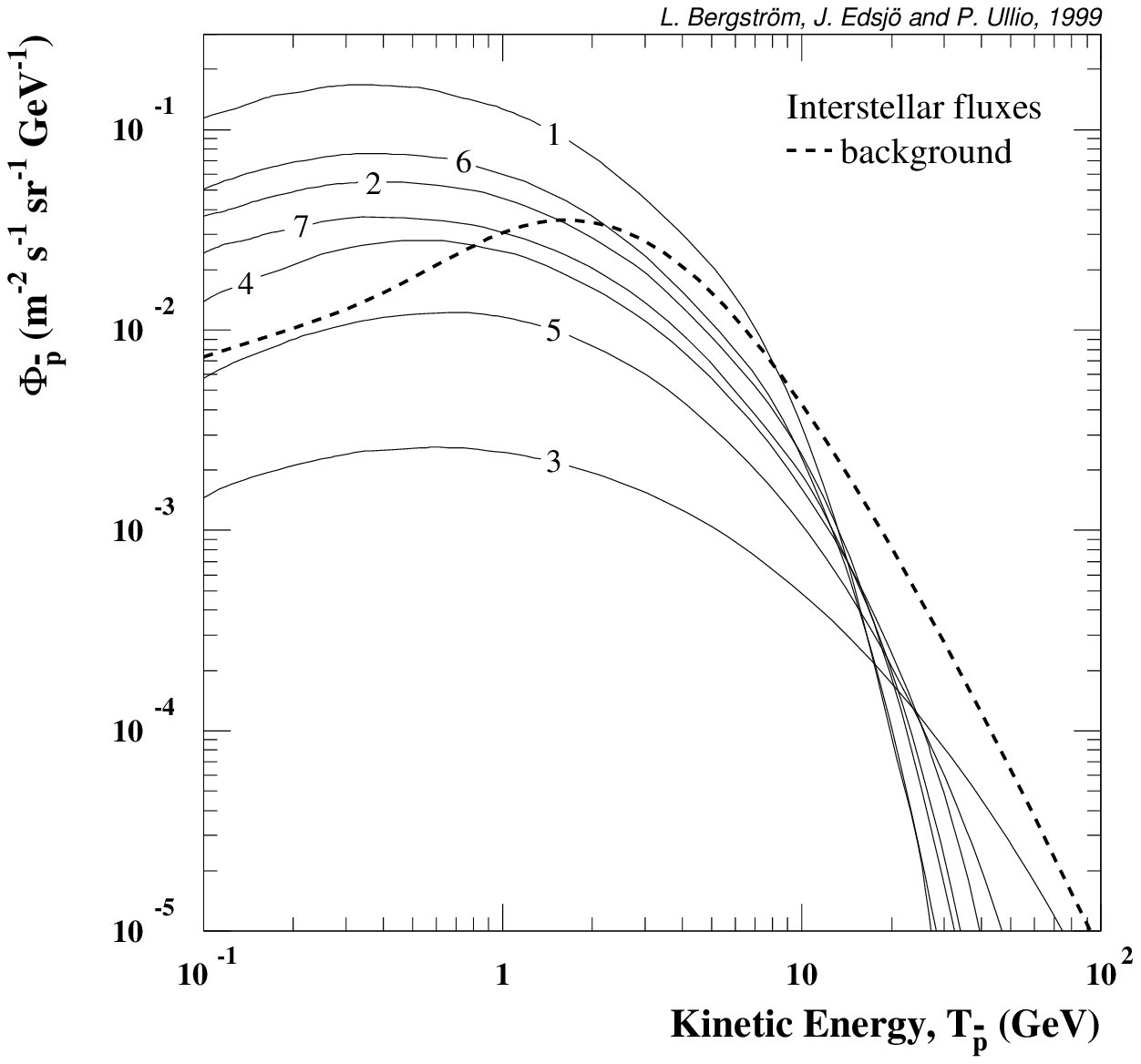,width=0.33\textwidth}}
\caption{\incap (a) The solar modulated antiproton fluxes at 0.35 GeV 
compared with {\sc Bess} 97. The models have been coded according
to their relic density, $\Omega_\chi h^2$. In (c) 
we show the flux of
antiprotons from neutralino annihilation at the optimal kinetic
energy, $T_{\rm opt}$, versus $T_{\rm opt}$. $T_{\rm opt}$ is defined
as the energy at which $\Phi_{\rm signal}/\Phi_{\rm background}$ is
highest and if the spectrum has more than one optimum, the highest two
have been included in the plot. The models have been coded according
to the neutralino mass in GeV\@. In (c) we show the 
antiproton spectra for 7 example models.}
\label{fig:pbsm}
\end{figure}

In Fig.~\ref{fig:pbsm} (a) we show the solar modulated fluxes versus 
the neutralino mass.  We see that there are many models with fluxes 
above the {\sc Bess} measurements.  However, this conclusion depends 
strongly on which range one allows for the neutralino relic density.  
In Fig.~\ref{fig:pbsm} (a) we have coded the symbols according to the 
relic density interval.  As can be seen, essentially all models which 
are in the {\sc Bess} measurement band have a relic density 
$\Omega_\chi h^2 < 0.1$.  If we instead require $0.1 \lsim 
\Omega_{\chi} h^2 \lsim 0.2$ the rates are never higher than the 
measured flux.

\begin{wrapfigure}[31]{r}{0.4\textwidth}
\epsfig{file=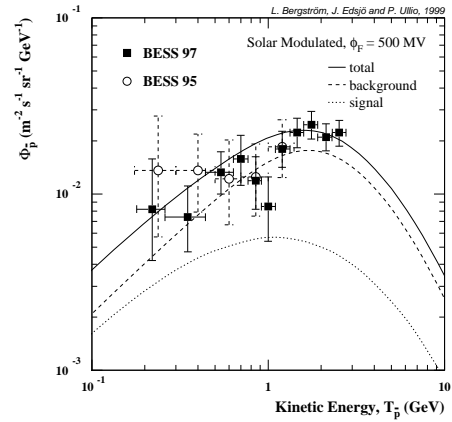,width=0.39\textwidth}
\caption{\incap An example of a composite spectrum consisting of our
  reference background $\bar p$ flux (Fig.\,\ref{fig:back} (b))
  reduced by 24 \% with the addition of the predicted flux from
  annihilating dark matter neutralinos of MSSM model number 5.}
\label{fig:pbsignal}
\end{wrapfigure}
This points to a weakness of this indirect method of detecting
supersymmetric dark matter: once the predicted rate is lower than the
presently measured flux, the sensitivity to an exotic component is
lost. This is because of the lack of a distinct signature which could
differentiate between the signal and the background. 

We are now interested in finding out if there are any special features
of the antiproton spectra from neutralino annihilation which
distinguish these spectra from the background. 
We then ask ourselves if there is an optimal energy at which
$\Phi_{\rm signal} / \Phi_{\rm background}$ has a maximum.
In Fig.~\ref{fig:pbsm} (b) we show the interstellar flux at
these optimal energies, $T_{\rm opt}$, versus $T_{\rm opt}$. We
have two classes of models: one class which have highest
signal to noise below 0.5 GeV (i.e.\ inaccessible in the solar system
due to the solar modulation) and one which have highest signal to
noise at 10--30 GeV.  For this first class of models, we note that
there exists a proposal of an extra-solar space probe
(Wells et al., 1998) which would avoid the solar modulation problem and
is thus an attractive possibility for this field. However, these
models have high rates in the range 0.5--1 GeV as well, even though it
would be even more advantageous to go to lower energies.  The second
class of models are much less affected by solar modulation and also
give reasonably high fluxes. 


In Fig.~\ref{fig:pbsm} (c) we show some examples of spectra.  They
show maxima occurring at lower energies than for our canonical
background.  At higher energies, the trend is that the slope of the
flux decreases as the neutralino mass increases.  Model number 3
corresponds to a heavy neutralino and its spectrum is significantly
less steep than the background.  If such a spectrum is enhanced, for
instance by changing the dark matter density distribution, we would
get a bump in the spectrum above 10 GeV (Ullio, 1999).

Finally, in Fig.~\ref{fig:pbsignal} we show an example of a
hypothetical composite spectrum which consists of our canonical
background flux decreased by 24 \% (obtained e.g.\ by decreasing the
primary proton flux by $1\sigma$), and the signal for model 5 in
Fig~\ref{fig:pbsm} (c). We can obtain a nice fit to the {\sc Bess}
data, but as noted before, there are no special features in the
spectrum that allow us to distinguish between this case and the case
of no signal.

\section{Discussion and conclusions}

We have seen that there is room, but no need, for a signal in the
measured antiproton fluxes. We have also seen that the optimal energy
to look for when searching for antiprotons is either below the solar
modulation cut-off or at higher energies than currently
measured. However, there are no special spectral features in the
signal spectra compared to the background, unless the signal is
enhanced and one looks at higher energies (above 10 GeV)\@.

We have stressed the somewhat disappointing fact that since the
present measurements by the {\sc Bess} collaboration already exclude a
much higher $\bar p$ flux at low energies than what is predicted
through standard cosmic-ray production processes, an exotic signal
could be drowned in this background. Even if it is not, the similar
shape of signal and background spectra will make it extremely hard to
claim an exotic detection even with a precision measurement, given the
large uncertainties in the predicted background flux (at least a
factor of a few, up to ten in a conservative approach). 

\section*{Acknowledgements}

We thank Mirko Boezio, Alessandro Bottino and collaborators, Per
Carlson and Tom Gaisser for useful discussions, Paolo Gondolo for
collaboration on many of the numerical routines used in the
supersymmetry part and Markku J\"a\"askel\"ainen for discussions at an
early stage of this project.  L.B. was supported by the Swedish
Natural Science Research Council (NFR).


\vspace{1ex}
\begin{center}
{\Large\bf References}
\end{center}
%
\setlength{\baselineskip}{0.9\baselineskip}
%
Adriani, O. \& al., 1995. Proc.\ of 24th ICRC,
Rome, {3}, 591.\\
%
Ahlen, S., \& al., AMS Collaboration. 1994,
Nucl. Instrum. Meth., A350, 351.\\
%
Alam, M.S.\ \& al. 1995,
({\sc Cleo} Collaboration), Phys. Rev. Lett., 74, 2885.\\
%
Ammar, R.\ \& al. 1993, ({\sc Cleo} Collaboration),
Phys. Rev. Lett., 71, 674.\\
%
Berezinskii, V.S., Bulanov, S.,
Dogiel, V., Ginzburg, V.\ \& Ptuskin, V. 1990, {\em Astrophysics
of cosmic rays}, North-Holland, Amsterdam.\\
Bergstr{\"o}m, L., Edsj{\"o}, J.\ \& Ullio, P.\ 1999, astro-ph/9902012.\\
%
Bottino, A., Donato, F.,
Fornengo, N. \& Salati, P. 1998, Phys.\ Rev.\ {D58} 123503.\\
%
Carr, J.\ 1998, 
The {\sc ALEPH} Collaboration. 1998, Talk by Carr, J., March 31, 1998,\newline
\href{http://alephwww.cern.ch/ALPUB/seminar/carrlepc98/index.html}
{http://alephwww.cern.ch/ALPUB/seminar/carrlepc98/index.html};
Preprint ALEPH 98-029, 1998 winter conferences,
\href{http://alephwww.cern.ch/ALPUB/oldconf/oldconf.html}
{http://alephwww.cern.ch/ALPUB/oldconf/oldconf.html}.\\
%
Chardonnet, P., Mignola, G., Salati, P.\ \&
Taillet, R. 1996, Phys.\ Lett., {B384}, 161.\\
%
Edsj{\"o}, J. 1997, PhD Thesis, Uppsala University, hep-ph/9704384.\\
%
Edsj{\"o}, J.\ \& Gondolo, P. 1997,
Phys.\ Rev., {D56}, 1879.\\
%
Gaisser, T.K. 1990, {\em Cosmic rays and
particle physics}, Cambridge University Press, Cambridge.\\
%
Ginzburg, V.L.\ \&  Syrovatskii, S.I. 1964, {\em The origin
of cosmic rays}, Pergamon Press, London.\\
%
Gleeson, L.J.\ \&
Axford, W.I. 1967, ApJ, {149}, L115.\\
%
Gleeson, L.J.\ \&
Axford, W.I. 1968, ApJ, 154, 1011.\\
%
Jungman, G.,
Kamionkowski M. \& Griest, K. 1996, Phys.\ Rep., 267, 195.\\
%
Matsunaga, H.\ \& al. 1998,
  Phys. Rev. Lett., {81}, 4052.\\
%
Orito, S. 1998, Talk given at The 29th
International Conference
on High-Energy Physics, Vancouver, 1998.\\
%
Ullio, P.\ 1999, astro-ph/9904086.\\
%
Wells, J.D., Moiseev, A.\ 
\& Ormes, J.F. 1998, preprint CERN-TH/98-362 (astro-ph/9811325).\\
~\\[-2ex]
\centerline{\it For a more detailed list of references, see
Bergstr{\"o}m, L., Edsj{\"o}, J.\ \& Ullio, P., 1999}

\end{document}